\documentclass{article}[11pt]

\usepackage[utf8]{inputenc}
\usepackage{array}
\usepackage{booktabs}
\usepackage[round, sort, numbers]{natbib}\setcitestyle{square}
\usepackage{csquotes}
\usepackage{graphicx}
\usepackage[left=1in,right=1in,top=1in,bottom=1in]{geometry}       
\usepackage{listings,verbatim}
\usepackage{tikz,pgfplotstable}\usetikzlibrary{positioning}\pgfplotsset{compat=1.11}
\usepackage{color,xcolor}             
\usepackage{latexsym,marvosym}
\usepackage{amssymb,amsmath,theorem,bm}
\usepackage{epsfig,subfig,float}
\usepackage{pdflscape}
\usepackage{tabularx}
\usepackage[bottom]{footmisc}
\usepackage{multirow}
\usepackage[hyphens]{url}
\usepackage[compact]{titlesec}

\newcommand\spacingset[1]{
    \renewcommand{\baselinestretch}{#1}\small\normalsize
}

\theoremstyle{plain}
\theoremheaderfont{\scshape}

\title{\bf \LARGE
    Anxiety, Alcohol, and Academics:
    A Textual Analysis of Student Facebook Confessions Pages \\ 
    (Research Note)\thanks{This research note is an extension of the 
    article \textit{``The Statistics of Tufts Confessions''} published
    by the Tufts Independent Data Journal on February 28th, 2015  
    (\texttt{http://tuftsenigma.org/the-statistics-of-tufts-confessions/}).
    A first draft of this research note was archived under the title 
    \textit{``Analyzing Latent Topics in Student Confessions 
    Communities on Facebook''} on June 17, 2015. Last significant edits were 
    made under the title \textit{``Anxiety, Alcohol, and Academics: A Large-scale 
    Textual Analysis of Student Facebook Confessions''} on March 19, 2018. No 
    significant edits have been made since March 19, 2018. For drafts of 
    active research papers, please see the author's webpage.}
}
\author{
    \Large Soubhik Barari\thanks{ 
    Department of Computer Science, Tufts University. 
    URL: \texttt{http://soubhikbarari.github.io/research.} }
}
\date{\Large First Draft: June 17, 2015 \\ This Draft: \today}

\begin{document}
\maketitle

\renewcommand{\abstractname}{\normalsize \bf Abstract}

\begin{abstract}
\normalsize
\noindent
What do college students reveal to their peers on social media under complete 
anonymity? Do their campus environments relate to the topics of their 
disclosure? To answer these questions, I analyze Facebook confessions pages. 
Popular on hundreds of college campuses, these pages allow students to 
anonymously post personal confessions on a public community forum. In this 
preliminary research note, I analyze several explanatory factors of online 
student confessional behavior. Aggregating nearly 200,000 confessions posts 
spanning a period of 3 years, I combine Latent Dirichlet Allocation (LDA) with 
human verification through Mechanical Turk to 
scalably identify topics in these online confessions. Where possible, I also 
link posts to real-world news events parsed from Twitter. I find that 
confessions mentioning socioeconomics as well as mental and physical health 
occur more often at top-ranking, expensive private colleges. 
While event-related confessions most often mention timely school-related 
events, many mention global and domestic events outside
of the local campus sphere. Results suggest that undergraduates from different 
campuses disclose about topics such as race, socioeonomics, and politics 
differently, but in aggregate, post in similar patterns over time. 
Additionally, results confirm that anonymous Facebook confessors receive 
support for confessions on important, but taboo topics such as health and 
socioeconomic status.
\end{abstract}

\break



\spacingset{1.5}

\section{Introduction}\label{sec:intro}

Facebook is a near ubiquitous platform where users, including adolescent-age 
college students, can express different dimensions of their identity including 
political ideology, sexual orientation, religion, and class status 
~\citep{pempek2009college}. Studies go further to show that Facebook usage can 
improve psychological well being and support self-esteem
~\citep{ellison2007benefits}. In particular, one's friend network can 
be a great resource: Facebook users often post statuses broadcasting requests 
for help or advice on various personal topics ~\citep{ellison2013calling}. 
However, these requests tend to be positive, steering clear of 
topics that might elicit controversy or embarassment such as depression, sex, 
or substance abuse ~\citep{kramer2011dimensions}. Where on social media might a 
college student be able to solicit advice or honestly disclose about these more 
taboo topics? 

In this research note, I examine Facebook Confessions Boards (FCBs), public 
forums used by university undergraduate communities to post anonymous personal 
confessions. These confessions are solicited through an open survey, filtered 
by a page administrator (typically a student in the community) and then posted 
anonymously on a public Facebook community page where page followers can 
comment and respond~\citep{simon2015reuters}. On a particular page, one 
anonymous post reads:

\spacingset{1}
\begin{displayquote}
\textit{``I honestly feel like a failure. I spent all semester looking for an 
internship, applied to dozens of places, got interviewed at only one and did 
not get it. If I cannot get even one, how will I be able to find a real job 
senior year?"}
\end{displayquote}
\spacingset{1.5}

\noindent Another confessor on yet another page writes:

\spacingset{1}
\begin{displayquote}
\textit{``I think I'm a Republican now, and I can't tell any of my friends or 
anyone else in [town], really."}
\end{displayquote}
\spacingset{1.5}

These posts describe personal secrets, controversies, and taboos; 
while topic and style varies, posts across pages follow a common theme of 
unfiltered disclosure. As in the second example message, confessors often 
acknowledge the difficulty of making such personal confessions to their own
social networks. Thus, FCBs provide a valuable anonymous channel for 
college students to express themselves in ways they cannot otherwise.

My note contributes and improves upon the extant research on anonymous social 
media in the following ways. First, I create a large-scale dataset by using 
computational and manual methods to scalably, but reliably code 173,218 
confessions posts (nearly double the dataset of previous studies). Second, I 
collect student life variables corresponding to the campus pages for our 
analysis. Third, I compare different campus pages to better 
understand user context and explore previously un-examined temporal 
and event-related dimensions of posting behavior. Finally, I provide a more 
thorough explanation of response behavior to posts by incorporating post topic, 
event mention, post language, and campus type in our analysis.

This note is structured as follows. I begin with a discussion of related work 
to contextualize my own research hypotheses. I then outline my methodology 
for data collection and analysis. Finally, I present the results of my 
analyses, discuss the general trends I find, and conclude.

\section{Background}\label{sec:background}

Studies articulate the stressful experiences of contemporary 
American college life, from underage binge drinking ~\citep{wechsler1994health} 
to socioeconomic identity struggle ~\citep{aries2007role} to mental 
health problems~\citep{hunt2010mental}. While common student issues persist 
\textit{across} many different university communities, evidence suggests that 
issues like racial and socioeconomic identity conflict differ in 
character \textit{between} different university communities. For instance, 
students at public and elite universities express different beliefs about the 
importance of socioeconomic status to their academic success 
~\citep{aries2007role}. Moreover, academic outcomes may differ between racially 
diverse and racially homogeneous institutions ~\citep{chang1999does}. Recent 
literature has focused on the detrimental effects of \textit{stigma} on 
racially and economically marginalized students' outcomes in elite and 
predominantly white universities respectively 
~\citep{johnson2011middle, pinel2005getting}. Additional work shows that race, 
class, gender, and other factors play important roles in 
generating personal stigma around mental health help-seeking 
behavior ~\citep{eisenberg2009stigma}.

Self-disclosure on Facebook can help to cope with such stresses
~\citep{wang2016modeling}. Ellison et al. examine how Facebook users 
access emotional support by broadcasting requests for advice via status updates 
to their network ~\citep{ellison2013calling}. However, users are unlikely to 
use public channels to share embarrassing or graphic content pertaining to 
issues and stigmas. Most English-speaking users tend to only post 
positive status updates ~\citep{kramer2011dimensions}. In contrast, anonymous 
online self-disclosure activates a \textit{disinhibition effect} allowing 
users to communicate personal issues and stigmas that they otherwise would not 
on public channels ~\citep{green2016social,joinson2007self}. Likewise, content 
perceived as controversial is more likely to be shared anonymously on social 
media~\citep{zhang2014anonymity}.

Several studies explicitly link college life with Facebook uage, 
however they are limited~\citep{ellison2007benefits, valenzuela2009there}. 
Ellison et al.'s 2007 study demonstrates a connection between Facebook usage 
and psychological well-being for undergraduates through the maintenance and 
formation of social capital formation \citep{ellison2007benefits} . However,
their study relies on self-reporting only examines public posting behavior. 
Birnoltz et al. (2015) specifically studies the anonymous question-asking and 
response-seeking behavior of undergraduates on Facebook confessions 
boards~\citep{birnholtz2015weird}. Analysis of 15,157 
question-asking posts in a 90,329 post dataset revealed a strong positive 
community response to questions on various stigmatized topics including mental 
health, sex, finances, and drugs. While Birnholtz et al. shows that male users 
engage more in response than female users, little else is known about the 
heterogeneity of community pages and users. 

In contrast, the larger corpus of university confessions presented in this note 
can measure the heterogeneity of posting behavior across campus 
communities. Below, I discuss these in the form of 
research questions.

\subsection{Institutional Differences}

Previous work has established how differences in academic institution along 
variables of public status, elitism, and campus diversity can produce both 
positive and negative effects for its students. I also observe these 
`institutional' effects in social media discourse, such as in the following 
anonymous confession:

\spacingset{1}
  \begin{displayquote}
  \textit{
``Of course, it is admirable to see talented individuals make it into 
outstanding institutions and also the fact that I've been accepted into one of 
those institutions is great and all, however I wished my parents wouldn't look 
at me like I'm a failure every time they looked at the `admissions to colleges' 
presentation on the school screen just because I didn't make it into the 
`top-[tiered]' colleges in the state.''}
  \end{displayquote}
\spacingset{1.5}

\noindent As such, we are interested in quantifying how the offline campus
context many predict the topic and language of online discourse: \\[-2ex]

\textsc{\textbf{RQ1:}} Do discussions of relevant confessional topics 
vary across different campus environments? \\[-2ex]

\subsection{Temporal Patterns}

University students have different stresses and concerns at 
different points during the academic year. Final exams are likely to be the 
prime academic stressor for students at the end of an academic term, while 
adjusting to a new social and academic climate is the more imminent concern 
at the start of a term. Furthermore, significant global news events as well as 
holidays are also likely to mobilize students to social media (e.g. major 
sports events, terrorist attacks). I hypothesize that students' online 
confessional behavior reflects these temporal events. Anecdotally, we find this 
to be true. One student in May of 2016 posts: \textit{``I'll go ahead and 
confess that finals are hard but I'm not worried anymore because next week I'm 
chilling on a beach in Juarez not a worry in the world''.} Another student in 
September of 2014 writes: \textit{``It's the start of school and I'm constantly 
checking the buzz only to find out when and where I can get some free candy and 
condoms.''} As such, we are interested in more broadly examining how posts 
differ across time and specifically in response to both 
locally and globally salient events: \\[-2ex]

\textsc{\textbf{RQ2:}} How do confessions change over time? How do 
confessing students respond to globally \textit{and} locally salient events?

\subsection{Audience Response}

\noindent Although likes and comments are notably sparse on 
these anonymous posts, a small minority of posts enjoy a large number of likes 
and comments. I find the highest number of likes and comments on any single 
post to be $24$ and $25$ respectively. Systematically, we are interested in 
factors \textit{across} pages that determine audience engagement with 
confessions: \\[-2ex]

\textsc{\textbf{RQ3:}} What factors explain audience responses to 
students' posts?

\section{Method}

Our university dataset consists of 2014 metadata from the U.S. National Center 
on Education Statistics. For all American Universities, I collect `social' 
variables of interest including undergraduate size, racial makeup, retention 
rate, and tuition price. Several categorical variables such as public status of 
the institution, religious affiliation, and campus setting were also used in 
our dataset. In addition, I aggregate the 2014 U.S. News ranking for each 
university. The U.S. News Rank in its calculation, for each school, bundles 
together components such as peer-determined academic reputation, student 
retention rate, faculty resources, student selectivity, and financial resources 
(per-student spending). The overall rank uses a weighted sum of each component 
score and normalizes all scores to a scale on 100. U.S. News Ranking is 
clearly highly collinear with other metrics in our dataset, however it provides 
an objective proxy measure for `elite-ness'. 

I then use our university dataset to query the Facebook Graph API for all 
affiliated confessions pages and their public posts. I parse a total of 195 
confessions pages (both active and non-active) and 173,218 posts dating from 
January 2013 to May 2016. For each post I collected the posting date as well as total likes and comments. 

Facebook confessions pages are widely geographically distributed -- states with 
the highest number of both active and non-active pages : California (72), New 
York (69), Illinois (67), Pennsylvania (67), and Massachusetts (52). However, 
pruning for only active pages ($>30$ posts), I reduce the number of pages to 
98 with the states with most pages being Minnesota (8), New York (7), 
Massachusetts (7), and Pennsylvania (7). The campuses with the highest number of 
parsed confession posts are Tufts University (27,605), Purdue University 
(10,629), University of Toledo (8,686), Biola University 
(8,288), Cornell University (7,845) and Montana State University (6,845).

In this note, I do not access any private or personal information about 
individuals, nor do I attempt to attain any data not already made accessible 
by Facebook's public API. All posts are completely anonymized by default. 
Furthermore, I refrain from identifying any institutions by name.

\subsection{Human-in-the-loop topic modelling}

To identify topics in the corpus of confessions, I deploy a 
``human-in-the-loop'' pipeline , which offers both scalability in automated 
topic annotation and reliability with human coder inputs.

I first deploy LDA (Latent Dirichlet Allocation) over a sample of documents in 
our corpus (N=27,000) in order to detect groups of keywords identifying 
possible topics of interest. LDA is a 
probabilistic method that can be used to find latent topics in textual 
data and their associated keywords. As such, this algorithm allows researchers 
to quickly discover co-occurrences of words and bin them together as topics in 
the text. Previous work in social media analysis has demonstrated LDA's 
effectiveness in analyzing social media corpora 
at scale without the need for human content coding~\citep{wang2013gender}. 
However due to its probabilistic nature, LDA can converge on vastly different 
topic-word sets on the same text each time it is deployed; for this reason, LDA 
can be an unstable and unreliable single identifier for topics in a 
text~\citep{koltcov2014latent}. 

Instead of relying on a single LDA run's results for our analysis, I run 
multiple configurations of the LDA algorithm over different model parameters. 
I then pick the top topic keywords from a number of topic-word groupings 
across different runs and manually construct a set of human interpretable 
topic-word sets with labels based on keyword similarity, familiarity of 
confessional content, as well as LDA topics previously observed in Birnholtz et 
al.'s subsample of confessions. In addition to stand-alone keywords for 
each topic set, I manually add commonly observed bigrams, trigrams, and
abbreviations. These are topically related phrases, euphemisms, expressions, 
and slangs that recurred across all campus pages e.g. 
``check your privilege" (n=65), used in reference to social privilege; ``sjw" 
(n=308), a shortening of `social justice warrior`; ``people of color" (n=98); 
``jerked off" (n=37). Using these keyword sets, each post in our corpus is 
coded as a topic if it contains the most words in the topic's associated 
keyword set. 

Previous work shows that strong human agreement with LDA-inferred labels can
further validate model-based content analysis approach ~\citep{dinakar2012you}. 
To evaluate my approach's robustness and, I use Amazon's Mechanical Turk -- 
shown to be an effective platform for crowdsourcing topic names 
~\citep{lau2011automatic} -- to collect 3 independent human codings of each 
post in a sample of 500 confessions as belonging to one of the chosen topic 
labels or irrelevant until $>80\%$ agreement was reached. The overall baseline 
agreement between 3 human coders on topics was 84\% (Fleiss's Kappa: 0.723), 
while overall agreement between the automated labeler and the 3 
human coders was comparably 80\% (Fleiss's Kappa: 0.66). I then use my 
final topics to categorize all posts in our dataset.

\subsection{Event labelling via Twitter}

In order to analyze how students respond to temporal events, I create an 
extensive compilation of headline news events via Twitter. From 2013 to 2015, 
I collect all relevant hashtags, named headlines, and descriptions of top 
trending global events from Twitter's `Year on Twitter' timelines. These 
include both events that occurred within a fixed time window and centered 
around a discrete incident (e.g. March Madness basketball championship, Boston 
Marathon bombings) and more vaguely defined events centered around movements 
(e.g. Black Lives Matter). I then create a similar dataset of significant 
university temporal events (e.g. graduation, matriculation, spring break, 
winter break).

To simplify the process of classifying posts relating to temporal events, I 
label a post as responding to an event if it explicitly mentions it by name or 
using obviously related synonyms (e.g. \textit{`vacation'} in place of 
\textit{`break'}, \textit{`bombings'} instead of \textit{`bombing'}) and 
keywords (e.g. \textit{`commencement'} -- a particular ceremony part of the 
larger \textit{`graduation'} event). To map global news events to related 
online social media terms, for each Twitter event, two independent coders were 
asked to construct a set of synonymous phrases they would use to 
explicitly refer to it on a social media forum. I constructed the final set of 
synonymous reference phrases for each event by taking the overlap of each 
coders' set of phrases and the phrases contributed by the papers' authors. 

I perform linguistic analysis of post content with the psycholinguistic 
lexicon LIWC ~\citep{pennebaker2001linguistic}. Hailed as the gold standard for 
linguistic analysis, LIWC parses a body of text and counts unigrams that 
reflect different cognitive states. LIWC is extensively used in 
applications of social media content analysis, such as predicting depression on 
Twitter ~\citep{de2013predicting}.

In the following section, I discuss the specific analyses conducted on the 
final dataset, the results and their broader implications.

\section{Results}\label{sec:results}

\subsection{RQ1}

First, we examine \textit{within-} and \textit{between-} page trends
in confessional topic and language. 

Out of a total of 173,218 posts in our dataset, I 
categorize 20,455 posts (12\%) into topics of interest. The proportions of 
topic occurrence resemble the frequencies of related taboo subjects in 
Birnholtz et. al's study -- the `Romantic/sexual' topic occurs most commonly, 
while `Academic' discussion is generally seen to be much less common. 
Table~\ref{fig:lda} displays the results of our topic labeling ordered by total 
volume of posts. 

\newcolumntype{C}{>{\centering\arraybackslash \small}p{60mm}}
\begin{table*}[h]
    \centering
    \resizebox{\textwidth}{!}{ 
        \begin{filecontents*}{topics.tsv}
topic	N	percent of posts	sample keywords
Romantic/sexual	12964	71.18	sexiest,hook up,girlfriend,in a relationship,lesbian,bisexual,raped,blowjob
Mental/physical health	2887	15.85	bipolar,depression,mental health,therapist,hospital,flu shot
Socioeconomic	723	3.97	expenses,money,poor,rich,middle class,wealth,student debt
Drugs/alcohol/partying	566	3.10	frat party,drunk,smokers,drugs,get high,alcohol,marijuana,acid
Political	443	2.43	feminism,staunch liberal,activists,politics,republican
Academic	381	2.09	pre med,lecture,transferring,in a class,gpa,failing
Race/ethnicity	248	1.36	people of color,issues of race,white privilege,ethnic,hispanic,asian
\end{filecontents*}
\pgfplotstabletypeset[
            col sep=tab,
            string type,
            columns/topic/.style={column name=\textbf{Topic}, column type={r}},
            columns/N/.style={column name=\textbf{N}, column type={c}},
            columns/percent of posts/.style={column name=\textbf{\%}, column type={c}},       
            columns/sample keywords/.style={
                column name=\textbf{Sample keywords}, 
                column type={l},
            },
            every head row/.style={before row=\hline,after row=\hline},
            every last row/.style={after row=\hline},
            ]{topics.tsv}
    }
    \spacingset{1}
    \caption{{\bf Discovered Topics and Sample of Associated Keywords}. This table 
    highlights each topic group, its name, and the top keywords discovered in the corpus 
    of all confessions. A total of 20,455 posts were labelled this way. Note that \% is 
    out of total number of labelled posts i.e. those not discarded as `Other'.}
    \label{fig:lda}    
\end{table*}

\noindent Are there are any \textit{within-page} trends in the discussion of 
certain topics? We find a strong negative correlation ($R^{2}=-0.86$) between 
the occurrence of `Romantic/sexual' topics and `Mental/physical health' in confessions pages. In 
other words, communities that discussed taboo sexual topics in high volume 
posted much less on issues of health and vice versa. Users of anonymous social 
media sites typically post about ``needs, wants, and wishes'' 
~\citep{correa2015many}, and the prevalence of different topics across campus 
pages likely reflect differences in these desires. Additionally, messages 
categorized as `Race/ethnicity' were found to have the 
highest average word count of $129(\pm 12)$, while `Drugs/alcohol/partying' 
posts were the most concise at an average word count of $59(\pm 4)$.

We now ask: how do topics of confessions vary \textit{between} pages? To answer 
this, I examine different binary groupings of colleges based on campus 
characteristics. For instance, to examine campus diversity, I compare `very 
white' colleges -- those with at least a standard deviation higher proportion 
of white students than the mean (\textgreater80.9\%) -- and `not very white' -- 
those less than a standard deviation away from the mean -- (\textless45.3\%). 
Additionally, I partitioned all pages into top ranking colleges (those with a 
significantly higher U.S. rank \textless20) vs. bottom ranking colleges (U.S. 
rank \textgreater170).

Next, for each pair of school groups, I cross-tabulate the number of topic 
posts originating from each group of pages and performed Chi-square analysis to 
examine if there were statistically different topic distributions between them. 
I find statistical differences between topic counts in both the diversity 
groups ($\chi^{2}=127.3$, df=6, $p=0.001$) and the rank groups 
($\chi^{2}=292.8$, df=6, $p=0.001$). Because I perform multiple comparisons, 
I apply Bonferroni correction when determining differences in individual topic 
counts.

\begin{table*}[h]
    \centering
    \resizebox{0.85\textwidth}{!}{ 
        \begin{tabular}{@{}rcccc@{}}
            \toprule
            & \multicolumn{2}{c}{\textbf{Campus diversity}} & \multicolumn{2}{c}{\textbf{U.S. News Rank}} \\
            \cmidrule(l{.5em}r{.5em}){2-3}\cmidrule(l{.5em}r{.5em}){4-5}
            \textbf{Topic of post}  & \shortstack{Very white \\ (\textgreater80.9\%)} & \shortstack{Not very white \\ (\textless45.3\%)} & \shortstack{Top \\ (\textless20)} & \shortstack{Bottom \\ (\textgreater170)} \\
            \midrule
            Academic & 19 (37) & 73 (54) & 68 (56) & 48 (59) \\
            Drugs/alcohol/partying  & 78 (57) & 63 (83) & 53 (63) & 76 (65) \\
            Mental/physical health  & 249 (284) & 447 (411) & 403 (362) & 338 (378) \\
            Political & 5 (16) & 35 (23) & 33 (18)* & 5 (19) \\
            Race/ethnicity & 2 (8) & 19 (12) & 19 (12) & 1708 (1518) \\
            Romantic/sexual & \textbf{1182 (1083)***} & 1472 (1570) & 1264 (1453) & \textbf{1708 (1518)***} \\
            Socioeconomic & 22 (69) & \textbf{149 (101)***} & \textbf{286 (158)***} & 36 (166) \\
          \midrule[\heavyrulewidth]
          \midrule[\heavyrulewidth]
            \multicolumn{5}{l}{Notes: \textsuperscript{*}$p<10^{-4}$, 
            \textsuperscript{**}$p<10^{-5}$, \textsuperscript{***}$p<10^{-6}$, expected 
            count in parantheses.} \\ 
        \end{tabular}
    }
    \spacingset{1}
    \caption{{\bf Cross-tabulation of Campus Type with Topic Occurrence.} This table 
    shows the results of multiple Chi-square tests on topic counts between different 
    categories of university pages. Bonferroni corrected $p$-values for individual count 
    comparisons are shown. }
    \label{fig:chisq}        
\end{table*}

From the results of our Chi-square analysis as shown in 
Table~\ref{fig:chisq}, we see that `Romantic/sexual', in particular, has the 
most disproportionate occurrences in the `Very white' and `Bottom ranking' 
cohort pages respectively. Conversely, `Socioeconomic' related posts occur more 
often in the `Not very white' and `Top ranking' cohort pages respectively. 
Additionally, `Political' themes are addressed more in the top ranking cohort 
of colleges, while `Academic' subject matter is discussed more in the `Not very 
white' confessional cohort.

Is there a clear relationship between campus environment and campus 
discourse? Using logistic regression, I measure the effects of specific campus 
characteristics on topic occurrences. When selecting predictors for our logistic 
regression model of post topic, I find U.S. News ranking to be highly correlated with 
full time student retention rate ($R^{2}=-0.9$), average net price 
($R^{2}=-0.78$), student to teacher ratio ($R^{2}=0.78$), and proportion of 
white students ($R^{2}=0.78$). Hence, I drop it as a covariate in the logit 
model. Chi-square tests between all pairings of categorical variables -- e.g. 
public status, religious affiliation -- revealed each one to be independent 
($p<0.005$), thus categorical variables of interest are kept as covariates in 
the logistic regression model.

\begin{table*}[!ht]
    \centering
    \resizebox{\textwidth}{!}{ 
    \begin{tabular}{@{}rccccccc@{}}

        \toprule

        \rule{0pt}{5ex}

          & \multicolumn{2}{c}{} & \multicolumn{3}{c}{\textbf{ \huge Dependent variable (topic of post)} } \\[0.15cm]

          \cmidrule(l{1em}r{1em}){2-8}

            {\huge \textbf{Predictors}}  
          & {\huge {Academic}} 
          & \shortstack{\huge {Drugs/alcohol}\\ \huge {/partying}} 
          & \shortstack{\huge {Mental}\\ \huge {/physical health}} 
          & {\huge {Political}} 
          & \shortstack{\huge {Race}\\ \huge {/ethnicity}} 
          & \shortstack{\huge {Romantic}\\\huge {/sexual}} 
          & \shortstack{\huge {Socio-}\\ \huge {economic}} \\[0.2cm]
        \midrule
        \shortstack[r]{\\ \thinspace \huge  \\ \huge Student body size \\ \huge (std. error) \\ \huge odds ratio } 
          & \shortstack{ \huge \textbf{0.003***} \\ \huge (0.0008) \\ \huge 1.003 \\ \thinspace } 
          & \shortstack{ \huge \textbf{0.020**} \\ \huge (0.006) \\ \huge 1.020 \\ \thinspace } 
          & \shortstack{ \huge \textbf{-0.006} \\ \huge (0.003) \\ \huge 0.994 \\ \thinspace } 
          & \shortstack{ \huge \textbf{-0.015} \\ \huge (0.009) \\ \huge 0.985 \\ \thinspace } 
          & \shortstack{ \huge \textbf{-0.001} \\ \huge (0.001) \\ \huge 0.999 \\ \thinspace } 
          & \shortstack{ \huge \textbf{0.003} \\ \huge (0.002) \\ \huge 1.003  \\ \thinspace } 
          & \shortstack{ \huge \textbf{-0.003***} \\ \huge (0.0007) \\ \huge 0.997 \\ \thinspace } 
          \\
      \shortstack{\huge Religious affil. \\ \thinspace \\ \thinspace \\ \thinspace \\ \thinspace \\ \thinspace \\ \thinspace \\ \thinspace \\ \thinspace \\ \thinspace  \\ \thinspace \\ \thinspace  } 
          & \shortstack{ \thinspace \\ \thinspace \\ \thinspace \\ \huge \textbf{-0.008} \\ \huge (0.125) \\ \huge 0.992 } 
          & \shortstack{ \thinspace \\ \thinspace \\ \thinspace \\ \huge \textbf{-0.140} \\ \huge (0.10) \\ \huge 0.869 } 
          & \shortstack{ \thinspace \\ \thinspace \\ \thinspace \\ \huge \textbf{0.044} \\ \huge (0.042) \\ \huge 1.04 } 
          & \shortstack{ \thinspace \\ \thinspace \\ \thinspace \\ \huge \textbf{-0.850***} \\ \huge (0.12) \\ \huge 0.427 } 
          & \shortstack{ \thinspace \\ \thinspace \\ \thinspace \\ \huge \textbf{-0.560***} \\ \huge (0.144) \\ \huge 0.571 }
          & \shortstack{ \thinspace \\ \thinspace \\ \thinspace \\ \huge \textbf{0.275***} \\ \huge (0.039) \\ \huge 1.316 } 
          & \shortstack{ \thinspace \\ \thinspace \\ \thinspace \\ \huge \textbf{-1.020***} \\ \huge (0.125) \\ \huge 0.361 } \\
      \shortstack{\huge Public \\ \thinspace \\ \thinspace \\ \thinspace \\ \thinspace \\ \thinspace \\ \thinspace \\ \thinspace \\ \thinspace \\ \thinspace \\ \thinspace \\ \thinspace \\ \thinspace \\ \thinspace } 
          & \shortstack{ \thinspace \\ \thinspace \\ \thinspace \\ \huge\textbf{-0.397} \\ \huge(0.22) \\ \huge 0.672 } 
          & \shortstack{ \thinspace \\ \thinspace \\ \thinspace \\ \huge\textbf{-0.520**} \\ \huge(0.173) \\ \huge 0.595 } 
          & \shortstack{ \thinspace \\ \thinspace \\ \thinspace \\ \huge\textbf{-0.210*} \\ \huge(0.082) \\ \huge 0.811 } 
          & \shortstack{ \thinspace \\ \thinspace \\ \thinspace \\ \huge\textbf{-1.929***} \\ \huge(0.0323) \\ \huge 0.145 } 
          & \shortstack{ \thinspace \\ \thinspace \\ \thinspace \\ \huge\textbf{-1.460***} \\ \huge(0.328) \\ \huge 0.232 } 
          & \shortstack{ \thinspace \\ \thinspace \\ \thinspace \\ \huge\textbf{0.583***} \\ \huge(0.068) \\ \huge 1.791 } 
          & \shortstack{ \thinspace \\ \thinspace \\ \thinspace \\ \huge\textbf{-0.910***} \\ \huge(0.186) \\ \huge 0.402 } 
          \\
      \shortstack{\huge Avg. net price \\ \thinspace \\ \thinspace \\ \thinspace \\ \thinspace \\ \thinspace \\ \thinspace \\ \thinspace \\ \thinspace \\ \thinspace \\ \thinspace \\ \thinspace \\ \thinspace  } 
          & \shortstack{ \thinspace \\ \thinspace \\ \thinspace \\ \huge\textbf{0.0006} \\ \huge(0.001) \\ \huge 1.00 } 
          & \shortstack{ \thinspace \\ \thinspace \\ \thinspace \\ \huge\textbf{-0.005} \\ \huge(0.010) \\ \huge 0.995 }  
          & \shortstack{ \thinspace \\ \thinspace \\ \thinspace \\ \huge\textbf{-0.002***} \\ \huge(0.0004) \\ \huge 0.998 }  
          & \shortstack{ \thinspace \\ \thinspace \\ \thinspace \\ \huge\textbf{0.007***} \\ \huge(0.001) \\ \huge 1.001 }  
          & \shortstack{ \thinspace \\ \thinspace \\ \thinspace \\ \huge\textbf{0.054***} \\ \huge(0.01) \\ \huge 1.055 }  
          & \shortstack{ \thinspace \\ \thinspace \\ \thinspace \\ \huge\textbf{0.002***} \\ \huge(0.0004) \\ \huge 1.002 }  
          & \shortstack{ \thinspace \\ \thinspace \\ \thinspace \\ \huge\textbf{-0.062***} \\ \huge(0.009) \\ \huge 0.940 } 
          \\
      \shortstack{\huge Retention rate \\ \thinspace \\ \thinspace \\ \thinspace \\ \thinspace \\ \thinspace \\ \thinspace \\ \thinspace \\ \thinspace \\ \thinspace \\ \thinspace \\ \thinspace \\ \thinspace \\ \thinspace } 
          & \shortstack{ \thinspace \\ \thinspace \\ \thinspace \\ \huge\textbf{0.0} \\ \huge(0.003) \\ \huge 1.0 } 
          & \shortstack{ \thinspace \\ \thinspace \\ \thinspace \\ \huge\textbf{-0.004*} \\ \huge(0.001) \\ \huge 0.996 } 
          & \shortstack{ \thinspace \\ \thinspace \\ \thinspace \\ \huge\textbf{0.002} \\ \huge(0.001) \\ \huge 1.002 }  
          & \shortstack{ \thinspace \\ \thinspace \\ \thinspace \\ \huge\textbf{0.009***} \\ \huge(0.002) \\ \huge 1.009 }  
          & \shortstack{ \thinspace \\ \thinspace \\ \thinspace \\ \huge\textbf{-0.013***} \\ \huge(0.002) \\ \huge 0.987 } 
          & \shortstack{ \thinspace \\ \thinspace \\ \thinspace \\ \huge\textbf{-0.001} \\ \huge(0.001) \\ \huge 0.999 }  
          & \shortstack{ \thinspace \\ \thinspace \\ \thinspace \\ \huge\textbf{0.008**} \\ \huge(0.002) \\ \huge 1.008 } 
          \\
        \shortstack{\huge \% of White students \\ \thinspace \\ \thinspace \\ \thinspace \\ \thinspace \\ \thinspace \\ \thinspace \\ \thinspace \\ \thinspace \\ \thinspace \\ \thinspace \\ \thinspace \\ \thinspace \\ \thinspace } 
          & \shortstack{ \thinspace \\ \thinspace \\ \thinspace \\ \huge\textbf{-0.010*} \\ \huge(0.004) \\ \huge 0.990 } 
          & \shortstack{ \thinspace \\ \thinspace \\ \thinspace \\ \huge\textbf{0.025***} \\ \huge(0.003) \\ \huge 1.025 }
          & \shortstack{ \thinspace \\ \thinspace \\ \thinspace \\ \huge\textbf{-0.007***} \\ \huge(0.001) \\ \huge 0.993 }
          & \shortstack{ \thinspace \\ \thinspace \\ \thinspace \\ \huge\textbf{-0.002} \\ \huge(0.004) \\ \huge 0.998 } 
          & \shortstack{ \thinspace \\ \thinspace \\ \thinspace \\ \huge\textbf{-0.010*} \\ \huge(0.005) \\ \huge 0.990 } 
          & \shortstack{ \thinspace \\ \thinspace \\ \thinspace \\ \huge\textbf{0.006***} \\ \huge(0.001) \\ \huge 1.006 } 
          & \shortstack{ \thinspace \\ \thinspace \\ \thinspace \\ \huge\textbf{-0.02***} \\ \huge(0.004) \\ \huge 0.980 }
          \\[0.2cm]
        \midrule
        \shortstack{\huge Constant \\ \thinspace \\ \thinspace \\ \thinspace \\ \thinspace \\ \thinspace \\ \thinspace \\ \thinspace \\ \thinspace \\ \thinspace  \\ \thinspace \\ \thinspace \\ \thinspace \\ \thinspace } 
          & \shortstack{ \thinspace \\ \thinspace \\ \thinspace \\ \huge\textbf{-3.38***} \\ \huge(0.481) \\ \huge 0.034 } 
          & \shortstack{ \thinspace \\ \thinspace \\ \thinspace \\ \huge\textbf{-4.56***} \\ \huge(0.411) \\ \huge 0.010 }
          & \shortstack{ \thinspace \\ \thinspace \\ \thinspace \\ \huge\textbf{-0.75***} \\ \huge(0.185) \\ \huge 0.472 } 
          & \shortstack{ \thinspace \\ \thinspace \\ \thinspace \\ \huge\textbf{-5.59***} \\ \huge(0.484) \\ \huge 0.003 } 
          & \shortstack{ \thinspace \\ \thinspace \\ \thinspace \\ \huge\textbf{-3.35***} \\ \huge(0.544) \\ \huge 0.035 } 
          & \shortstack{ \thinspace \\ \thinspace \\ \thinspace \\ \huge\textbf{0.04} \\ \huge(0.153) \\ \huge 1.041 } 
          & \shortstack{ \thinspace \\ \thinspace \\ \thinspace \\ \huge\textbf{-0.70} \\ \huge(0.390) \\ \huge 0.496 } 
          \\
        \shortstack{\thinspace \\ \huge Likelihood ratio test \\ \huge vs. null model $\chi^{2}(6)$ \\ \thinspace } 
          & \shortstack{{\huge \textbf{67.24***}} \\ \thinspace \\ \thinspace \\ \thinspace \\ \thinspace \\ \thinspace } 
          & \shortstack{{\huge \textbf{82.38***}} \\ \thinspace \\ \thinspace \\ \thinspace \\ \thinspace \\ \thinspace } 
          & \shortstack{{\huge \textbf{81.05***}} \\ \thinspace \\ \thinspace \\ \thinspace \\ \thinspace \\ \thinspace } 
          & \shortstack{{\huge \textbf{418.54***}} \\ \thinspace \\ \thinspace \\ \thinspace \\ \thinspace \\ \thinspace } 
          & \shortstack{{\huge \textbf{248.81***}} \\ \thinspace \\ \thinspace \\ \thinspace \\ \thinspace \\ \thinspace } 
          & \shortstack{{\huge \textbf{324.62***}} \\ \thinspace \\ \thinspace \\ \thinspace \\ \thinspace \\ \thinspace } 
          & \shortstack{{\huge \textbf{232.95***}} \\ \thinspace \\ \thinspace \\ \thinspace \\ \thinspace \\ \thinspace  } 
          \\[0.3cm]
        \midrule[\heavyrulewidth]
        \midrule[\heavyrulewidth]
        \multicolumn{8}{l}{\huge \shortstack{Notes: $N=26777$, \textsuperscript{*}$p<0.05$, \textsuperscript{**}$p<0.01$, \textsuperscript{***}$p<0.001$,  student body size quantified in thousands of students, \\ avg. net price quantified in thousands of U.S. dollars.}} \\ 
    \end{tabular}
    }
    \spacingset{1}
    \caption{{\bf Effects of Student Life Factors on Topic Occurrence.} This table 
    illustrates the results of a logistic regression of LDA topic occurrence on various 
    institutional variables.}
    \label{fig:logistic}        
\end{table*}

The results of the logit model are shown in Table~\ref{fig:logistic}, 
where we can note individual topic predictors with high odds ratios \footnote{A 
predictor's logistic coefficient is the effect of its unit increase on the log 
odds of a specific topic occurrence.}. In particular, note that campus pages of 
public institutions have nearly twice ($1.791$) the odds of a `Romantic/sexual' 
topic occurrence, and less than half the odds of a `Political' or `Race/
ethnicity' topic occurrence than non-public institutions. Interestingly, we see 
that campuses with religious affiliation have increased odds of `Romantic/
sexual' topic discourse and a similar decrease in the odds of `Socioeconomic' 
topic discourse. We also note that a higher average tuition increases the odds 
of `Race/ethnicity' mention and, surprisingly, decreases the odds of 
`Socioeconomic' mention. What is also unexpected is that decreased campus 
diversity (smaller proportion of white students) does not produce a huge effect 
on race discourse.

Turning to a finer linguistic analysis, I compare LIWC language 
characteristics between different campus groups. Using the Mann-Whitney U test, 
I find that pages in the top ranking cohort had a higher proportion 
($p < 0.001$) of \textit{anxiety} words (mean=0.548) than pages in the bottom 
ranking cohort (mean=0.294) in their respective messages. Similarly, I find 
that confessors of the `Not very white' campus cohort used more \textit{family} 
words ($p < 0.001$) in their posts about the `Mental health' topic (mean=0.278) 
than confessors of the `Very white' group (mean=0.174). `Not very white' 
cohort confessors similarly used more \textit{family} words ($p < 0.001$) in 
their `Socioeconomic' mentions (mean=0.878) than the `Very white' (mean=0.063). 
Meanwhile, we see that `Very white' cohort confessions contain more 
\textit{friend} keywords ($p < 0.01$) in reference to `Mental health' 
(mean=0.74) than `Not very white' confessions (mean=0.67). However, we see no 
meaningful linguistic differences between the `Very white' and `Not very white' 
cohorts in posts about `Race/ethnicity'. The mean word count for a `Race/
ethnicity' post from a `Not very white' campus page is $100(\pm26)$, while it 
is $160(\pm100)$ for `Very white' campus pages.

Can a school's attributes be predicted by its students' confessions?
With random sub-sampling cross-validation over 20 rounds, a random forest 
trained on page topic proportions across colleges was able to predict the 
whiteness group (`Very white', `Not very white') of an out-of-sample school 
with $84\%$ accuracy ($\pm 3\%$). All other institutional categories were found 
to have poor predictiveness with similar techniques.

\subsection{RQ2}
We turn our attention to temporal variations in Facebook confessions. 

\begin{figure}[h]
\centering
\includegraphics[width=0.9\textwidth]{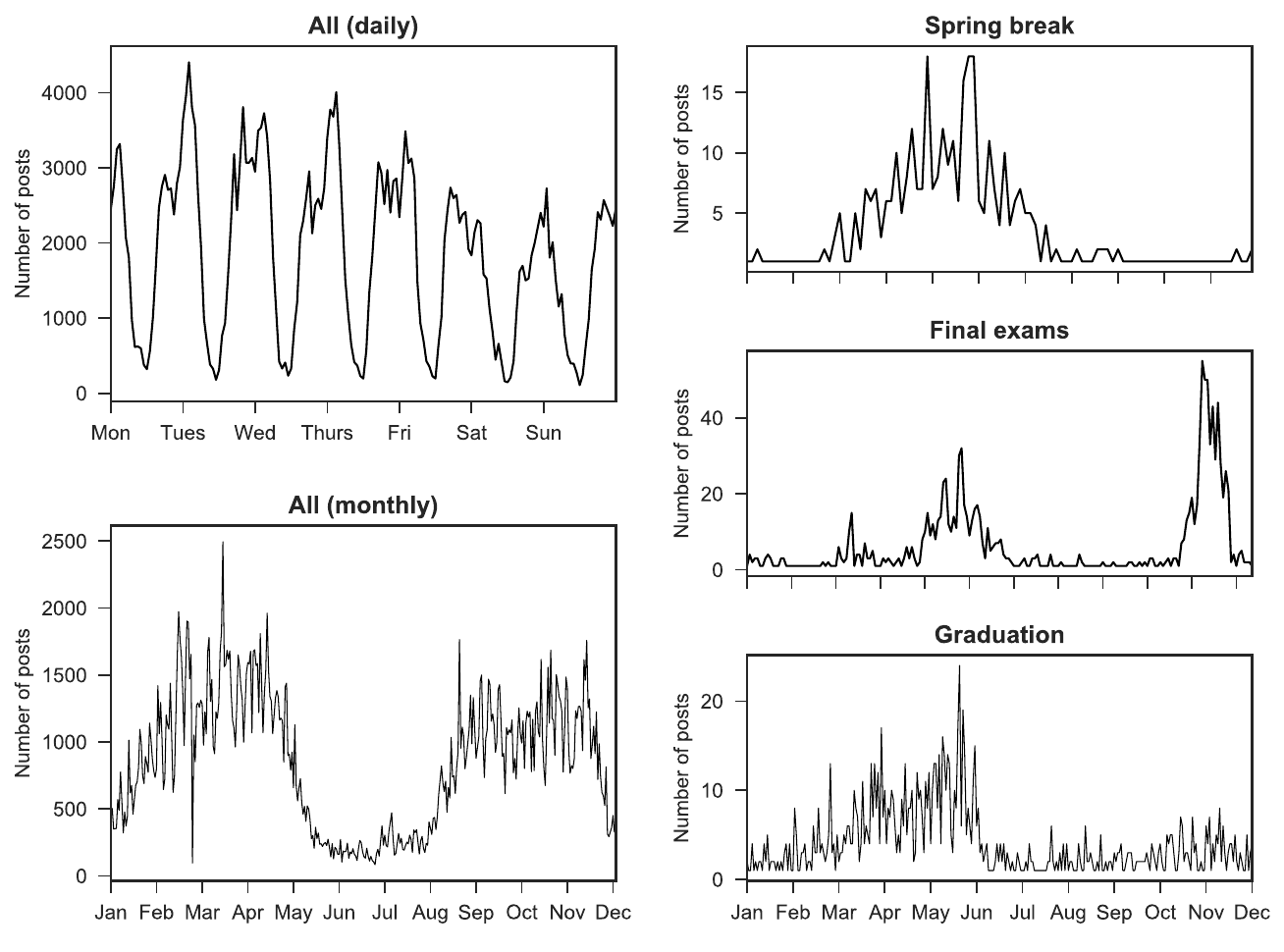}
\spacingset{1}
\vspace*{-0.5cm}\caption{
  {\bf Cumulative Volume of Confessions Posts by Category.} This figure shows all posts 
  extracted between January 2013 and May 2016 belonging to the selected
  temporal window(s) and event categories. 
}\label{fig:events} 
\end{figure}

Figure~\ref{fig:events} displays the temporal trends in confession posts. On 
the monthly scale, we observe spikes in volume of posts in mid-March, early 
May, September, mid-November, and December which roughly correspond to spring 
vacation, graduation, commencement, Thanksgiving vacation, and winter vacation 
respectively. On the weekday scale, we observe more posting activity at the 
start of the week, with a slight spike near the end of the academic week and an 
ebb during the weekend. Overall, we see a much larger body of confessions 
around university local events than global events, with `term-ending' school 
events having the most mention (e.g. graduation, final exams). 

\pagebreak

\begin{landscape}
\vspace*{\fill}
 \begin{table*}[!ht]
 \vspace{-0.65cm}
    \centering
    {\renewcommand{\arraystretch}{1.0}
    \resizebox{1.35\textheight}{!}{
        \begin{tabular}{@{}rccr@{}}
            \midrule
              { \textbf{Event}} 
            & { \textbf{Example post excerpt}} 
            & \shortstack{ \textbf{Number} \\ \textbf{of posts}}
            & \shortstack{ \textbf{Longest post streak} \\ \textsc{(streak start--end)}}
            \\
            \midrule
               {Graduation}
                  & \textit{``..Graduating as an engineer this spring and starting to 
                  get some interviews...''}
                  & 1225
                  &  48 \textsc{(Apr 9--May 28 2013)} \\

               {Final exams}
                  & \textit{``I had sex in a study room during finals week last year!''}
                  & 1224
                  & 37 \textsc{(Apr 17--May 27 2013)} \\

               {Spring break}
                  & \textit{``I think [university] should have an alternative spring 
                  break...''}
                  & 370
                  &  29 \textsc{(Mar 3--Apr 5 2013)} \\

               {Winter break}
                  & \textit{``...Coked out for the first time during winter break...''}
                  & 185
                  & 7 \textsc{(Dec 24--Jan 1 2014)} \\

               {Summer break}
                  & \textit{``I hate my family and wish they'd stop asking when my 
                  summer break starts...''}
                  & 53
                  & 3 \textsc{(May 12--May 16 2014)} \\

               {Matriculation}
                  & \textit{``...Post an `I fucking love college' status while you're at 
                  matriculation''}
                  & 4
                  & 3 \textsc{(Sep 15--Sep 18 2015)} \\

                  \midrule

               {New Years}
                  & \textit{``I can't wait to fucking get out of this room... New year 
                  hurry the hell up!''}
                  & 324
                  & 20 \textsc{(Dec 18--Jan 11 2015)} \\
  
               {Police Shootings}
                  & \textit{``The situation in Ferguson just breaks my heart.. for 
                  Michael Brown, his family..''}
                  & 151
                  & 16 \textsc{(Nov 25--Dec 12 2014)} \\

               {Super Bowl}
                  & \textit{``Anybody want to watch the Super Bowl together?''}
                  & 29
                  & 4 \textsc{(Feb 2--Feb 7 2015)} \\

               {2014 Ebola Outbreak}
                  & \textit{``Whenever I see someone sneeze or cough I'm convinced they 
                  have Ebola.''}
                  & 27
                  & 6 \textsc{(Sep 30--Oct 6 2014)} \\

               {2014 World Cup}
                  & \textit{``...the only one at [university] excited about 
                  the 2014 FIFA World Cup.''}
                  & 26
                  & 2 \textsc{(Nov 3--Nov 5 2015)} \\
              \addlinespace[0.1cm]
               {\shortstack[r]{2013 Boston Mara- \\thon 
              Bombings}}
                  & \shortstack{\textit{``So scared to hear the news of the Boston 
                  Marathon explosions..''} \\\quad }
                  & \shortstack{13 \\\quad \\\quad}
                  & \shortstack{2 \textsc{(Apr 11--Apr 13 2014)} \\\quad} \\ 

               {Nelson Mandela Death}
                  & \textit{``Here come all the tools who will start posting Mandela 
                  quotes...''}
                  & 13
                  & 2 \textsc{(May 19--May 21 2015)} \\

               {March Madness}
                  & \textit{``I indirectly blame March Madness for my current bout of 
                  depression.''}
                  & 12
                  & 2 \textsc{(Mar 21--Mar 25 2013)} \\

               {Robin Williams Death}
                  & \textit{``'On behalf of [university], I'd like to say we loved you, 
                  Robin Williams... ''}
                  & 12
                  & 2 \textsc{(Feb 11--Feb 13 2013)} \\

               {\#RefugeesWelcome}
                  & \textit{``I don't see the slightest sign of [Muslim students'] 
                  support for the refugees'' }
                  & 9
                  & 2 \textsc{(Sep 15--Sep 17 2015)} \\

               {Coachella}
                  & \textit{```I wish Coachella wasn't 
                  so goddamn expensive.'' }
                  & 8
                  & 2 \textsc{(Mar 13--Mar 15 2016)} \\

               {2015 Paris Attacks}
                  & \textit{``What happened in Paris is terrible, but ISIS
                  has already killed over 170,000 people...'' }
                  & 6
                  & 2 \textsc{(Dec 1--Dec 3 2015)} \\

               {Grammys}
                  & \textit{``Grammys suck this year.''}
                  & 5
                  & 2 \textsc{(Feb 7--Feb 9 2015)} \\
              \addlinespace[0.1cm]
               {\shortstack[r]{2014 Malala Yousafzai \\Nobel Prize}}
                  & \shortstack{\textit{``I am thrilled that Malala Yousafzai won the Nobel Peace Prize...''} \\\quad }
                  & \shortstack{5 \\\quad \\\quad}
                  & \shortstack{2 \textsc{(Nov 7--Nov 9 2014)} \\\quad} \\ 

            \midrule[\heavyrulewidth]
            \midrule[\heavyrulewidth]

        \end{tabular}}}
    \spacingset{1}
    \caption{
        {\bf Summary of Event-Related Posts.} This table gives a volume and `streak' 
        summary of selected high-volume and well-labeled events. A \textit{streak} is 
        defined as a window of days for which at least one confession contained a mention of the event. }
\label{fig:eventsAll}        
\end{table*}
\vspace*{\fill}
\end{landscape}

Posts across both types of events are often humorous or blunt, however, 
there is evidence that suggests strands of serious civic engagement and political 
engagement. Notably, we see an effort to post rhetoric and/or factual information 
regarding world crises or political events such as the 2015 Paris Attacks:

\spacingset{1}
\begin{displayquote}
    \textit{
        ``Murder is not the proper response to cartoons. No logical person could believe 
        this. If you honestly believe that radical Islamic terrorism is anything more 
        than a bastardization of the world's most widespread religion, or that twisting 
        the words of the Muslim prophets to support violent attacks on civilians is 
        justified, you are devoid of all human decency and morals.''
    }
\end{displayquote}
\spacingset{1.5}

\noindent Another post responding to the `\#BringBackOurGirls' movement (not 
displayed in Table~\ref{fig:eventsAll}):

\spacingset{1}
\begin{displayquote}
    \textit{
        ``I legitimately don't understand the intense scrutiny placed on Israel compared 
        to other world conflicts. I'm not saying that it should be ignored just because 
        there are worse things out there, but it seems to take a completely 
        disproportionate amount of criticism.''
    }
\end{displayquote}
\spacingset{1.5}

\noindent Similar posts can be found mentioning the Syrian refugee crisis. While 
Birnholtz et al. find positive comment responses to posts outcrying for help, I find 
that posts \textit{themselves} also offer solidarity, and resources in response to 
tragedy events. The full posts corresponding to the excerpts presented in 
Table~\ref{fig:eventsAll} for the events, `Robin Williams Death' and `2013 Boston 
Marathon Bombings' both posted phone numbers for suicide and mental health 
helplines respectively.

\begin{figure*}[!b]
  \centering
    \makebox[\textwidth][c]{
        \includegraphics[width=0.95\textwidth]
        {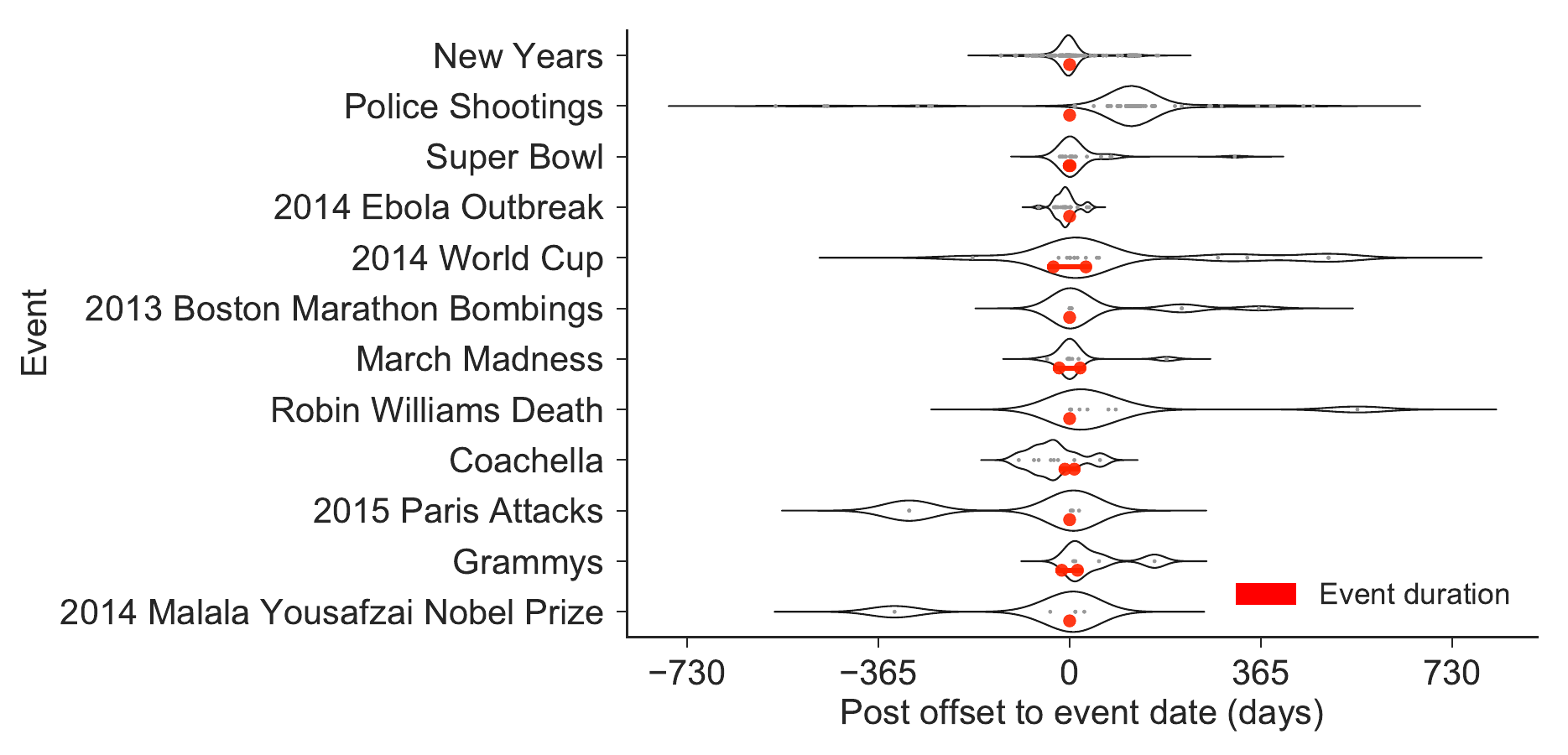} 
    }
    \spacingset{1}
    \vspace*{-0.65cm}
    \caption[]{
        {\bf Post Alignment with Domestic and World Events.} We considered `New 
        Years' an annual event and treated the offset of posts as the closest 
        difference to either New Years or New Years Eve. To best discretize the 
        event `Police Shootings', I looked at posts relevant to the 
        initial Ferguson shooting by labelling posts only containing Ferguson 
        related phrases in our phrase set. For the `2014 Ebola Outbreak', I 
        considered the discrete event of the WHO announcement of Ebola in the 
        U.S. (Oct. 10, 2014). Events like `Coachella' and `March Madness' are 
        considered annual events over a specific date range.
    }
\label{fig:eventDecay}   
\end{figure*}

Figure~\ref{fig:eventDecay} shows patterns of temporal alignment of posts 
mentioning global event with their respective events. Not surprisingly, I 
observe different posting patterns around different types of global events: 
anticipatory posts about an annual event (e.g. `Coachella'), reflective posts 
about events (e.g. `2014 World Cup', `Grammys'), a burst of concentrated posts 
around a discrete global phenomenon (e.g. `New Years', `2014 Ebola Outbreak'), 
and prolonged reactionary discourse as a single discrete event spills over into 
other related events (e.g. `Police Shootings'). Regarding local events, 
students there is a much wider spread of posts around `Spring break' than any 
other event; `Final exams' posts align with the end of spring and winter terms 
with a higher peak in winter, and `Graduation' posts sharply spike at the end 
of May. Although students do sparsely discuss global, non-campus-related 
events, it is evident that crises and tragedies mobilize students to online 
anonymous posting.

\begin{figure*}[!h]
  \centering
    \makebox[\textwidth][c]{
        \includegraphics[width=0.95\textwidth]{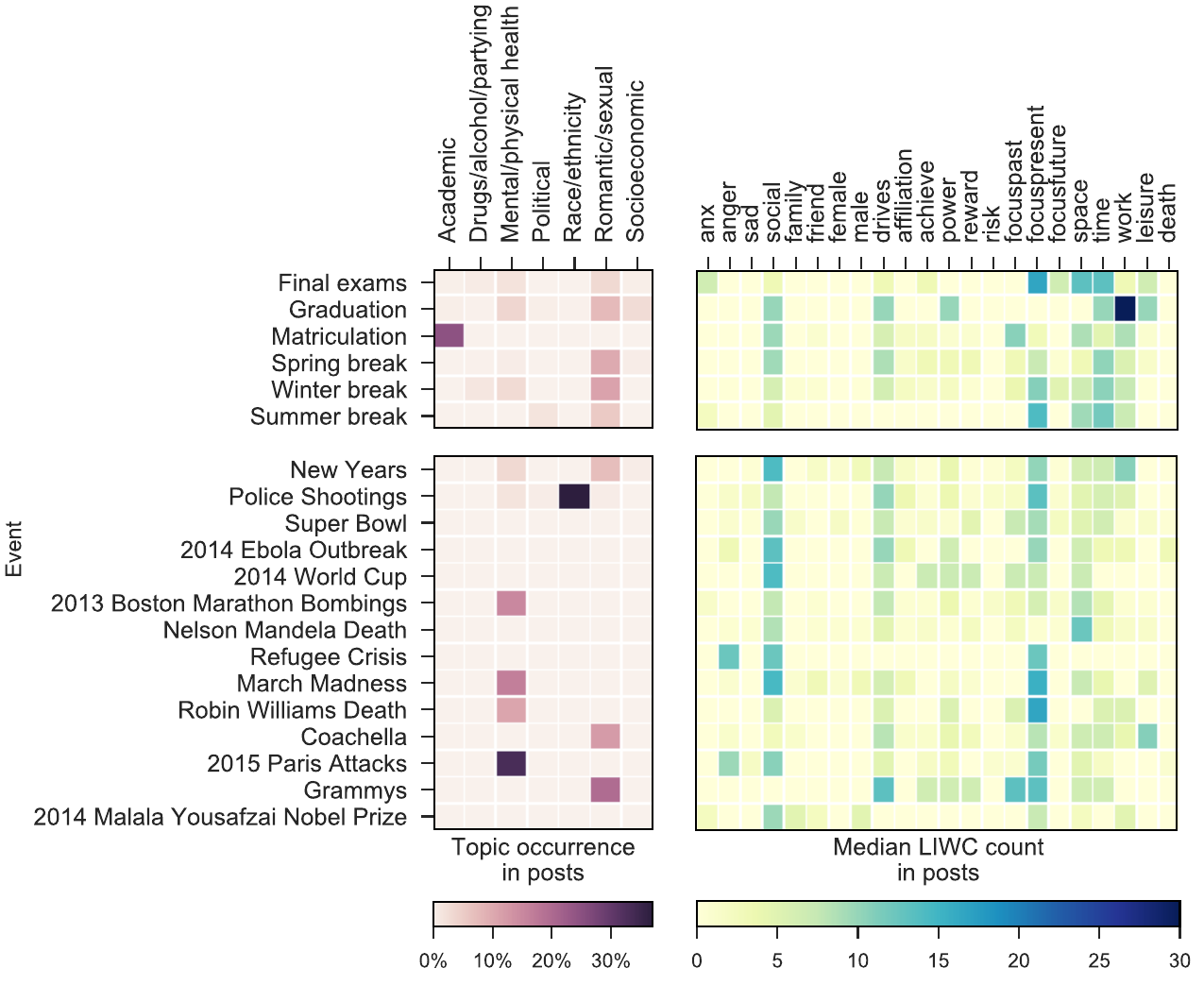} 
    }
    \vspace*{-0.65cm}

    \spacingset{1}
    \caption[]{
        {\bf Language Heatmap for Event-Related Posts.} This figure matches 
        posts corresponding to each `local' and `global' event with 
        corresponding counts of topics as well as median LIWC counts.
    }
    \label{fig:heatmap}   
\end{figure*}

How does posting language differ across different types of events? 
Figure~\ref{fig:heatmap} shows how language differs along topics and linguistic 
categories.

Generally speaking, local events co-occur with `Romantic/sexual' 
dialogue while global tragedy events initiate conversation about `Mental/
physical health' (with the exception of March Madness which where mental health 
may be brought up for ironic effect as shown in Table~\ref{fig:eventsAll}). 
Notably, `Police Shootings' is the only event that triggers `Race/ethnicity' 
discourse. The \textit{social} linguistic category is highly present across all 
global events, though generally less so in non-tragic, non-political events 
(`Grammys', `March Madness', `Coachella'). The \textit{focuspresent} category 
appears the most in the typical `Final exams' post and the least in the typical 
`Graduation' post; we observe \textit{work} category language around 
`Graduation'. In general, \textit{anger} is most found around politically 
salient global events. Overall, we find no statistical differences in how 
events are discussed on temporal or language dimensions \textit{across} 
different types of campuses. 

\subsection{RQ3}

In this section, we address the question of audience engagement to a post's 
topic as well as its page context. To measure this, I perform a zero-inflated 
negative binomial regression ~\citep{zuur2009zero} on post topic and 
institutional factors of the originating page with likes/comments as responses 
respectively.

I choose the zero-inflated model due to the large number (8,287) of posts with 
zero responses. In our case, the Negative Binomial specification is preferred 
to the Poission regression model due to the independence of the count variance 
from the mean, allowing for a more accurate modelling of counts. The resulting 
coefficients are shown in Table~\ref{fig:nb}. Substantively, we can interpret 
the coefficient values as the unit increase in like/comment count in response 
to a unit increase of the covariate. Additionally, to account for post response 
between confessions pages, I include a posts's originating page as a random 
effect in our model. Finally, we choose our zero-inflated mixed model approach 
as it scores a lower AIC than both the Poisson model and the fixed effect 
negative binomial model across all regressions.

\begin{figure*}[!ht]
    \centering
    \resizebox{0.65\textwidth}{!}{ 
    \begin{tabular}{@{}rcccccccc@{}}
        \toprule
        \textbf{Topic of post} & \textbf{Likes} & \textbf{Comments} \\
        \midrule
        \ Academic & \textbf{0.339(0.127)**} & 0.183(0.219) \\ 
        \ Drugs/alcohol/partying & \textbf{0.25(0.11)*} & 0.214(0.163) \\
        \ Mental/physical health & \textbf{0.338(0.05)***} & \textbf{0.340(0.064)***} \\
        \ Political & 0.19(0.12) & 0.080(0.144) \\
        \ Race/ethnicity & \textbf{0.561(0.173)**} & 0.122(0.281) \\
        \ Romantic/sexual & \textbf{-0.08(0.030)**} & \textbf{0.121(0.040)**} \\
        \ Socioeconomic & \textbf{0.272(0.141)*} & \textbf{0.716(0.186)***} \\
        \midrule
        \ Constant & \textbf{1.60(0.210)***} & 0.102(.188) \\ 

        \midrule[\heavyrulewidth]
        \multicolumn{2}{l}{ \shortstack{Notes: $N=26777$, \textsuperscript{*}$p<0.05$, \textsuperscript{**}$p<0.01$, \textsuperscript{***}$p<0.001$.}} \\ 
    \end{tabular}
    }
    \spacingset{1}
    \caption{{\bf Negative Binomial Regression of Post Responses on Topic.} 
    This table summarizes the results of a negative binomial regression of post response
    (comments/likes) on post topic.}
    \label{fig:nb}        
\end{figure*}

From Table~\ref{fig:nb}, we see that the occurrence of `Race/
ethnicity' and `Mental/physical health' have the highest effects on likes while 
`Mental/physical health' and `Socioeconomic' have the highest effects on 
comments. This confirms Birnholtz et al.'s conclusion that posts on finances 
are more likely to create conversation than other taboo topics. It is 
noteworthy that `Political' posts are significantly less likely to receive 
likes or comments relative to other topics.

\section{Discussion}

My results offer a detailed characterization of the anonymous online discourse 
between and within college campuses in the United States. Regarding top U.S. 
News ranking colleges, we find a significantly larger volume of `Socioeconomic' 
posts than about `Romantic/sexual'. However, we do not see evidence that more 
expensive colleges result in greater `Socioeconomic' student discourse. It is 
possible that posts about middle class marginalization at the most expensive 
universities are highly polarized, but fewer in proportion to 
other topics; moreover, such discourse may be more prevalent amongst the top 
tier of colleges, but not strictly monotonic with tuition expense within this 
tier. Linguistically, the presence of more \textit{anxiety} words further 
suggests that elite college confessions are more frequently concerned with 
serious, `restricted' personal subjects (e.g. `Mental/physical health', 
`Socioeconomic', `Political') rather than the less serious topics more often 
found in most other student confessions (e.g. `Drugs/alcohol/partying', 
`Romantic/sexual').

Regarding race on campus, we do not observe a strict positive relationship 
between student diversity and `Race/ethnicity' topic counts. However, the 
language of \textit{family} words may differentiate such 
campuses from their less diverse peers; students of minority or immigrant
status may be relating their taboo confessions more to their family than white
students. In general, confessional language 
is distinct between between campuses of vastly different student 
racial makeups, as evidenced by the strong predictability of campus diversity 
from page language traits. 

Taboo subjects receive more likes and comments across all campuses than 
lighthearted ones. Temporally, students are most likely to confess on these 
topics following the weekend (when posts recount anecdotes) or 
the end of an academic semester (when posts reflect on significant semester
experiences). In a broader temporal context, students post in 
relation to events around them -- university (local) events are mentioned in 
greater volume and longer sustained windows while global news events 
are generally discussed less frequently with varying degrees of immediate 
reactivity and/or anticipation. However, ubiquituous 
world/domestic events (e.g. New Years, Super Bowl, World Cup), and politically 
salient events (e.g. Police shootings, Ebola, marathon bombings) do draw 
relatively large numbers of prolonged confessions.

This study faces several important limitations. Notably, each 
post is constrained to only belong to a single topic, while the traditional LDA 
topic model posits that documents contain a mixture of topics. While my choice 
to discretize posts to the most strongly signaled topics enables us to 
confidently label a post's topic as relevant or not, we lose the opportunity to 
analyze potentially interesting collocations of topics (i.e. `Mental/physical
health' and `Socioeconomic'). Additionally, we were unable to extract any 
meaningful information from Facebook `reactions', a set of emotive responses 
available to Facebook users to disambiguate their response to shared content, 
due to the small sample size (`reactions' were introduced in February 2016) and 
over-dispersion of `reactions'.

In future work, I hope to gather more posts, in particular posts with 
`reactions'. Moreover, I encourage researchers to explore 
applications that may predict campus sentiment in response to 
university and real world events as well as causal mechanisms explaining the 
behavioral trends that I have found in this initial study.

\section{Conclusion}

In this research note, I analyze institutional, temporal, and audience 
response trends in Facebook confessions by university students in the United 
States. Using a total of 173,218 confessions, I find that students at 
top-ranking elite universities post more about socioeconomics and 
mental/physical health. However, I find no evidence that higher tuition 
price predicts more socioeconomic discourse or that lower student diversity 
predicts less race discourse. While university students most engage with 
local events (evidenced in both posting volume and posting duration), 
they do meaningfully post about world events and politically salient news. 
My results confirm that American college students across different 
campuses disclose private information in stylistically different ways, 
while in the aggregate, still post in similar patterns over time and 
receive support on taboo, but important matters such as mental health 
and socioeconomic status. 

\break
\spacingset{1}
\bibliographystyle{plainnat}
\bibliography{ref}

\nocite{*}

\end{document}